\documentclass[twocolumn]{aastex631}

\usepackage{amsmath}
\usepackage{bm}
\usepackage{graphicx}

\newcommand{\msol}{$M_\odot$~}
\newcommand{\ron}{$R_{1.4}$~}
\newcommand{\lone}{$\Lambda_{1.4}$~}

\begin{document}

\title{Model-Independent Determination of the Tidal Deformability of a 1.4 \msol Neutron Star from Gravitational-Wave Measurements}

\author[0000-0001-6406-1003]{Chun Huang}
\affil{Physics Department and McDonnell Center for the Space Sciences, Washington University in St. Louis; MO, 63130, USA}
\correspondingauthor{Chun Huang}
\email{chun.h@wustl.edu}


\begin{abstract}
Tidal deformability of a \(1.4\,M_\odot\) neutron star provides a pivotal window into the physics of dense nuclear matter, bridging gravitational-wave(GW), electromagnetic observations and nuclear physics. In this work, we present a novel, data-driven approach to constrain \(\Lambda_{1.4}\) without invoking specific equation-of-state(EOS) models. By interpolating directly over the mass--tidal-deformability posteriors from GW170817, we obtain an EOS-independent constraint of
$
\Lambda_{1.4} \;=\; 222.89_{-98.85}^{+420.33}.
$
We further combine these GW-based results with the X-ray EOS-independent constraint from ~\cite{Huang_2025}, deriving a multimessenger limit of
$
\Lambda_{1.4} \;=\; 265.18_{-104.38}^{+237.88},
$
which remains largely EOS agnostic. This framework demonstrates that higher-order terms neglected in linear expansion methods do not significantly affect \(\Lambda_{1.4}\) estimates under current observational uncertainties. As gravitational-wave detectors improve in sensitivity and more binary neutron-star mergers are discovered, our purely data-driven strategy can serve as a robust standard baseline for extracting neutron-star interior properties without relying on unverified EOS models.
\end{abstract}



\section{Introduction} \label{sec:intro}
A \(1.4\,M_{\odot}\) neutron star---often referred to as a canonical-mass neutron star---is deeply intertwined with various aspects of nuclear and astrophysical research \citep{Lattimer2001,Lattimer2013,Drischler21,lim24}. Its central density, approximately twice the nuclear saturation density, can be probed by ground-based nuclear experiments such as P-REX and C-REX \citep{crex, prex}. In gravitational-wave (GW) studies, this canonical mass also serves as the usual reference point for computing the tidal deformability \citep{PhysRevLett.111.071101,PhysRevD.92.023012}. Since the landmark detection of the first binary neutron-star merger, GW170817, by the LIGO/Virgo Collaboration\citep{Abbott_2017,Abbott_2018,Abbott_2019}, using tidal measurements to investigate neutron-star interiors has become a particularly active field eg. \citep{Malik_2018,2018ApJ...852L..29R,Raithel_2019,PhysRevLett.123.141101,De_2018,Most_2018,Dietrich_2020,Ghosh_2022,Huth_2022,Huang:2023grj,Huang:2024rvj}. One key parameter reported in these studies is \(\Lambda_{1.4}\), the tidal deformability of a \(1.4\,M_{\odot}\) neutron star. Extracting \(\Lambda_{1.4}\) from GW data, however, is nontrivial. In the case of GW170817, the total mass of the binary is \(2.73^{+0.04}_{-0.01}\,M_{\odot}\), with one component spanning \(1.36\text{--}1.62\,M_{\odot}\) and the other \(1.15\text{--}1.36\,M_{\odot}\). Both companions could plausibly be near \(1.4\,M_{\odot}\), yet neither is precisely \(1.4\,M_{\odot}\).

{The tidal deformability of a canonical $1.4\,M_\odot$ neutron star, $\Lambda_{1.4}$, has emerged as a important observable for neutron-star equation-of-state (EOS) research \citep{Hinderer_2008,Read_2009,Damour_2009,Abbott_2018}{.}  {Because $\Lambda\propto k_2 C^{-5}$, where $k_2$ is the second Love number and $C$ the stellar compactness, a precise determination of $\Lambda_{1.4}$ simultaneously encodes information about the stellar compactness and the underlying pressure of cold, $\beta$-equilibrated matter at densities of $\sim\!2$--$3$ times nuclear saturation \citep{2001ApJ...550..426L}.} Canonical-mass stars occupy the density regime that overlaps most directly with terrestrial constraints from chiral effective field theory and electroweak probes of finite nuclei, making $\Lambda_{1.4}$ an ideal quantity for bridging astrophysical and laboratory data \citep{Hebeler_2010,Drischler_2016,Tews_2018,De_2018}.  Moreover, many global EOS studies adopt $\Lambda_{1.4}$ as a primary calibration target because it is less sensitive to high-density extrapolations than, for example, the maximum mass, and it scales monotonically with the symmetry-energy slope parameter $L$, permitting direct comparison with neutron skin measurements. (see more in \cite{2001ApJ...550..426L,crex,prex}). Consequently, an EOS-agnostic constraint on $\Lambda_{1.4}$ provides a stringent benchmark for present and future theoretical models of neutron star interior.
}

A standard method in gravitational wave physics for estimating \(\Lambda_{1.4}\) relies on a linear expansion of \(\Lambda(m)\,m^5\) around \(m = 1.4\,M_{\odot}\), see \cite{PhysRevLett.111.071101,PhysRevD.92.023012}. While straightforward, neglecting higher-order terms means this approach is only an approximation. Another common strategy adopts a parametric EOS model---whether physically motivated or agnostic---to infer \(\Lambda_{1.4}\) . However, selecting a specific EOS framework can introduce considerable systematic uncertainties, because different EOS models often yield significantly different predictions for \(\Lambda_{1.4}\). (eg. \cite{Eemeli,kumar23,Raaijmakers_2019,Raaijmakers_2020,Raaijmakers_2021,Huang:2023grj,Huang:2024rvj,Rutherford_2024}). In addition, the exact EOS of neutron-star matter remains an open question in the field.

Motivated by these challenges, a variety of model-independent approaches have been proposed, including nonparametric inference schemes and advanced GW data-analysis frameworks that reduce reliance on explicit EOS assumptions \citep{Landry_2019,Essick_2020,Landry:2020vaw,Biswas_2022,Breschi_2024}. In particular, for GW170817, the fact that both stellar components lie so close to \(1.4\,M_{\odot}\) suggests that \(\Lambda_{1.4}\) might be extracted directly through analysis of the \((M,\Lambda)\) posteriors, thereby minimizing additional modeling assumptions. This work builds on a method previously applied to determine the radius of a \(1.4\,M_{\odot}\) neutron star by interpolating between two \((M,R)\) measurements lying near this canonical mass~\citep{Huang_2025}. Here, we adapt that technique to infer \(\Lambda_{1.4}\) for GW170817. Because the binary's component masses are naturally clustered around \(1.4\,M_{\odot}\), the same procedure can be employed to provide a more direct, potentially less model-dependent estimate of \(\Lambda_{1.4}\). This study therefore represents a natural extension of the original approach. 

After inferring \(\Lambda_{1.4}\) from the gravitational-wave side via our model-agnostic approach, we can incorporate the constraints from \cite{Huang_2025} to form a multi-messenger bound on this key parameter. This synthesis represents {a novel,} purely data-driven, model-independent constraint on neutron star interior properties, thereby avoiding the ambiguities {inherent to specific theoretical models for the equation of state,  such as piece-wise polytropes \citep{2009PhRvD..79l4032R,Hebeler_2013}, speed-of-sound parametrizations \citep{PhysRevC.98.045804,2019MNRAS.485.5363G}, chiral effective-field-theory extrapolations \citep{Drischler_2021,PhysRevC.88.025802}, and relativistic mean-field parameterizations \citep{10.1093/pasj/psu141,PhysRevC.90.055203,Huang:2023grj} etc that introduce model-dependent systematic uncertainties that are difficult to quantify.} With this result in hand, it becomes possible 
to compare astrophysical findings directly with related experimental outcomes in nuclear physics, providing a robust check on how well the two domains agree. 

{Recent advances in terrestrial nuclear physics further refine the value of  $\Lambda_{1.4}$ by  Parity-violating electron–scattering experiments such as PREX-II \citep{prex} and CREX \citep{crex}. Both of these experiments aim on measuring the neutron-skin thickness $\Delta R_{\mathrm{np}}$ of heavy nuclei, thereby constraining the density dependence of the symmetry energy through its slope parameter $L$ at saturation density.  Theoretically, $L$ sets the pressure of neutron-rich matter just above saturation and is therefore strongly correlated with the radius and tidal deformability of a canonical $1.4\,M_\odot$ neutron star \citep{Fattoyev_2018,PhysRevLett.127.232501}.  By providing a model-independent bound on $\Lambda_{1.4}$ that is free from assumptions about any specific EOS parametrisation, our analysis supplies an individual astrophysical result that can be compared directly with the $L$ values $R_{1.4}$ and $\Lambda_{1.4}$ inferred from PREX-II and CREX.  Such a cross-comparison offers a stringent, theory-independent test of whether nuclear interactions calibrated at near-saturation densities remain compatible with the macroscopic properties of neutron-star matter.
}

In the following sections, we first provide a concise overview of our methodology for analyzing tidal deformability measurements in Section~2, where we also introduce the relevant GW170817 dataset. Section~3 presents our inference results under two scenarios: one treating the two neutron stars as having distinct masses, and another assuming that one of the stars has exactly 1.4~\(\mathrm{M_{\odot}}\). In Section~4, we compare our findings with those from Ref.~\cite{Huang_2025}, enabling a multi-messenger constraint on \(R_{1.4}\) and \(\Lambda_{1.4}\). 
Section~5 contrasts these results with constraints derived from nuclear experiments and other astrophysical observations. Finally, in Section~6, we summarize our conclusions and discuss future implications of our work.

\section{Methodology and GW170817 Data Sets}
In this section, we briefly outline our statistical methodology, originally presented in \cite{Huang_2025}, and describe the revisions needed for treating tidal deformability measurements. We also clarify the underlying assumptions guiding our analysis. We apply this approach to three variants of the GW170817 dataset: the high-spin prior, low-spin prior, and EOS-insensitive scenarios. The first two are effectively independent of any specific equation of state (EOS), while the third is explicitly EOS-insensitive. Although our method can be applied to parametric EOS analyses, we choose not to include those datasets here in order to maintain a strictly model-independent framework.
\begin{figure}
  \centering
\includegraphics[width=\linewidth]{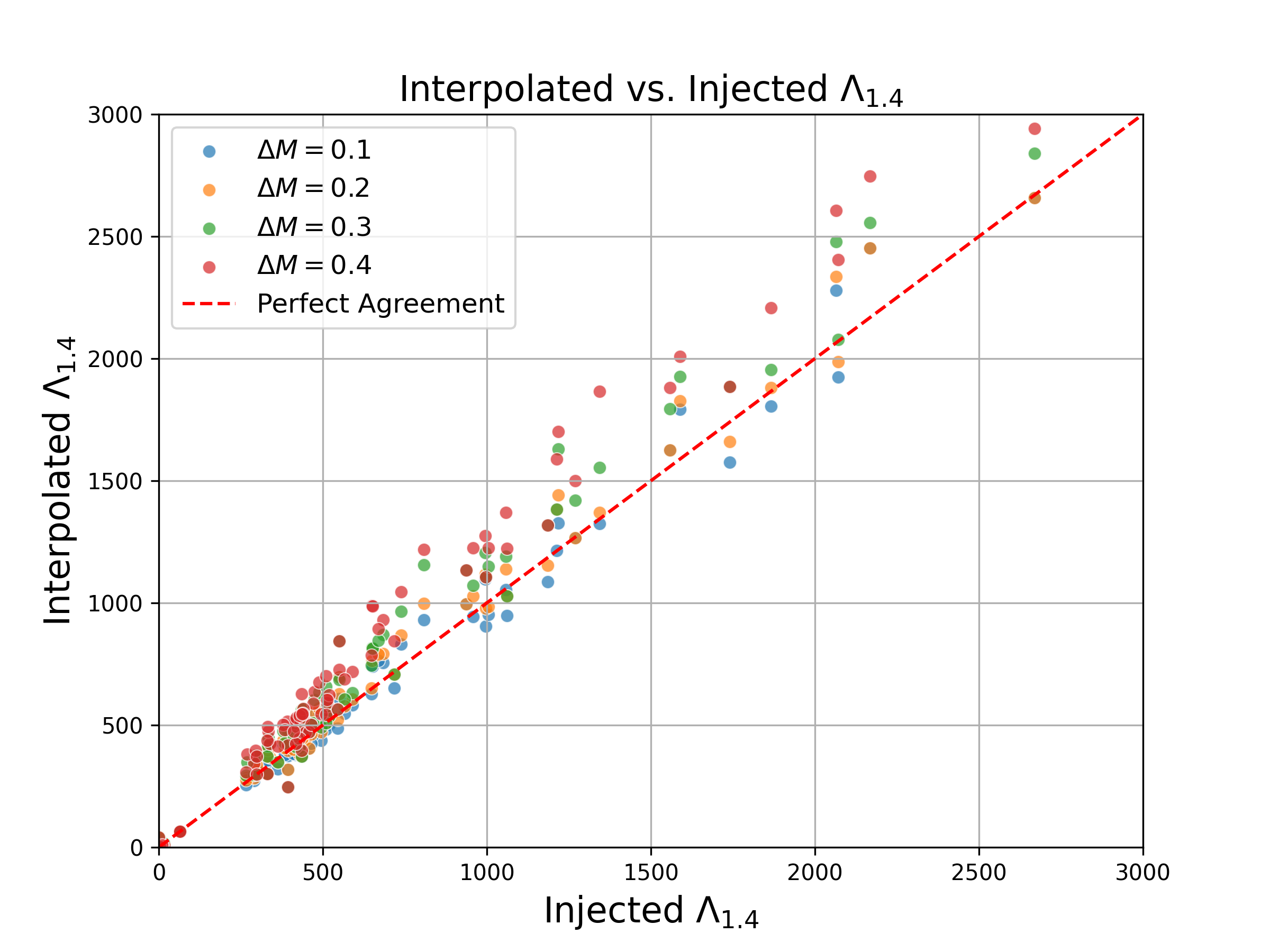}
  \caption{Comparison between $\lambda_{1.4}$ recovered by linear interpolation and the injected (‘true’) value for 200 synthetic binaries covering $\Delta M= 0.1–0.4\,M_{\odot}$  . Colours denote the mass gap; the dashed red line marks perfect agreement}
  \label{fig:bias_test}
\end{figure}
\begin{figure*}
    \centering
    \begin{minipage}{0.48\linewidth}
        \centering
        \includegraphics[width=1\linewidth]{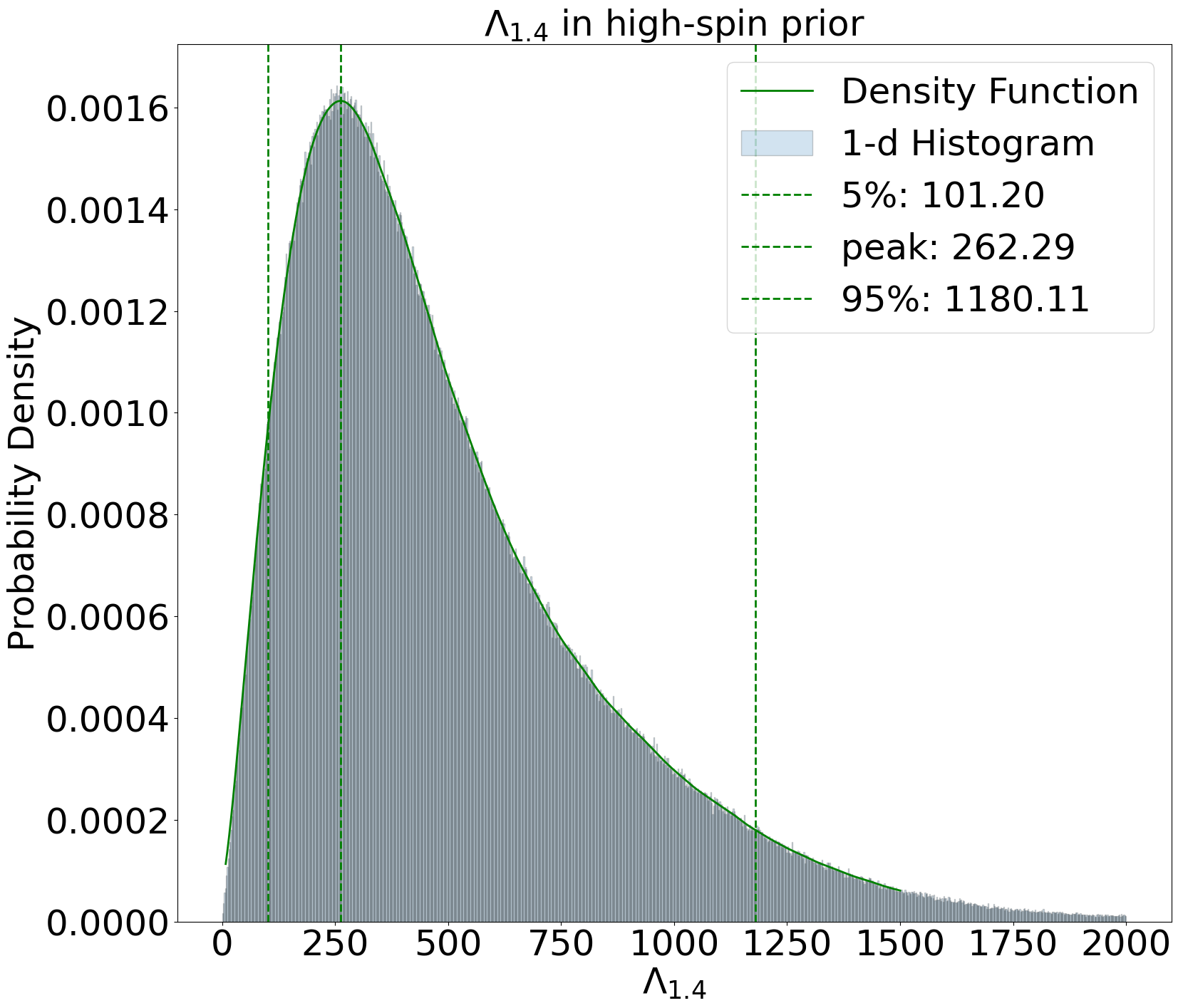}
        \label{highspin}
    \end{minipage}%
    \begin{minipage}{0.48\linewidth}
        \centering
        \includegraphics[width=1\linewidth]{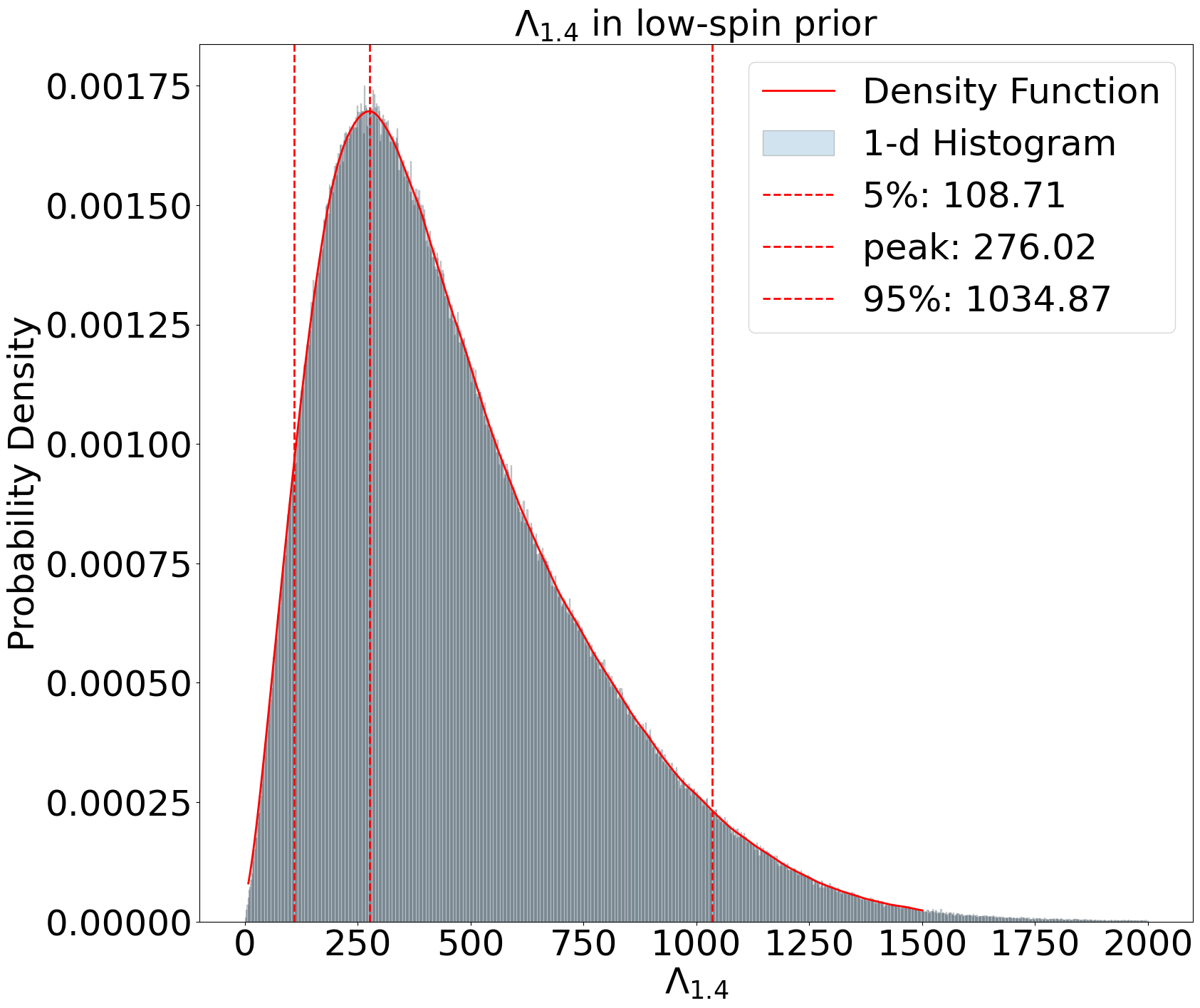}
        \label{lowspin}
    \end{minipage}%
    \caption{The 1-D distribution of 1.4~\msol{} star tidal deformablity \lone from high-spin prior data set (left) and low-spin prior data set (right): The dashed lines in each figure from left to right represent the quantiles of each distribution at 5\% quantile maximum probablity density peak value location and 90\% quantile. The density function is computed using KDE estimation.}
    \label{fig:highlow_compare}
\end{figure*}
\begin{figure}
	\centering
	\includegraphics[scale=0.3]{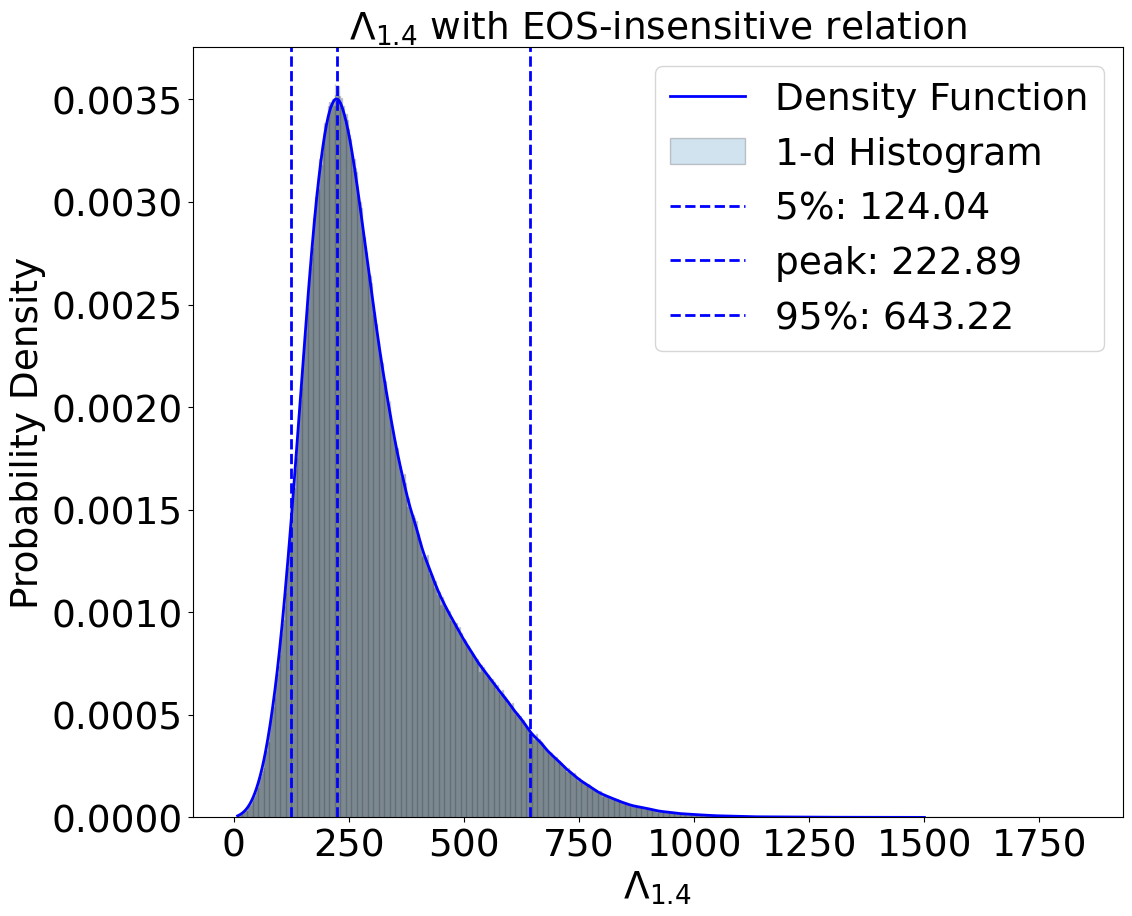}
	\caption{The 1-D distribution of 1.4~\msol{} star tidal deformablity \lone from EOS-insensitive  data set (right): The dashed lines in each figure from left to right represent the quantiles of each distribution at 5\% quantile maximum probablity density peak value location and 90\% quantile. The density function is computed using KDE estimation. 
	}
	\label{EOS_independent}
\end{figure} 
\subsection{Inference Methodology: Scenario 1}
As previously discussed in \cite{Huang_2025}, we investigate two complementary scenarios. The first scenario considers a binary system where the two neutron stars have strictly different masses and are not equal to 1.4 \msol. The second scenario examines the possibility that one of them have the same mass as 1.4 \msol.

In the first scenario, we focus on the mass--tidal-deformability measurements \((M_i, \Lambda_i)\), where \(i\) labels the individual stars. We approximate the posterior probability distribution of these parameters using a kernel density estimation (KDE) for both \(M_i\) and \(\Lambda_i\). From the resulting KDEs, we generate \(3 \times 10^6\) random samples of \((M_i, \Lambda_i)\), deliberately excluding cases in which the two stars share the same mass. This exclusion avoids discussion of twin-star configurations, which are beyond the scope of this work, and maintains consistency with the approach in \cite{Huang_2025}.

For the remaining samples, we select pairs such that
$
M_1 < 1.4\,M_\odot < M_2,
$
and then apply a linear interpolation between these pairs to estimate the tidal deformability \(\Lambda_{1.4}\) at a mass of \(1.4\,M_\odot\). Repeating this process over all qualifying pairs yields a distribution of \(\Lambda_{1.4}\) values. In the analysis of \cite{Abbott_2018}, \(\Lambda_1\) is assumed to be larger than \(\Lambda_2\) (i.e., the more massive star has a lower tidal deformability), reflecting the typical expectation for realistic EOSs. We adopt the same condition here. Indeed, imposing a negative slope in the \(M\text{--}\Lambda\) relation offers tighter constraints on \(\Lambda_{1.4}\) compared to our previous radius-based analysis in \cite{Huang_2025}, which did not require any slope condition in the \(M\text{--}R\) plane.

As described in \cite{Huang_2025}, this data-driven method avoids specifying any particular EOS model. {Our key assumption is that, within a narrow $\Delta M \approx 0.3M_\odot$ window centered on $1.4\,M_\odot$, the mapping $m\!\mapsto\!\Lambda(m)$ is \emph{empirically linear} for all viable EOSs considered to date.  The physical rationale is summarized in the following subsection 2.4
} In the specific case of GW170817, the mass difference between the two components is small enough that this linear assumption remains valid.

\subsection{Inference Methodology: Scenario 2}

In the second scenario, we supplement the first method by considering the possibility that one of the stars in the binary system have the same mass of \(1.4\,M_\odot\). Because one component in GW170817 has a mass distribution indicating an upper bound of \(1.36\,M_\odot\), we do not combine the information here as in \cite{Huang_2025}. Instead, we seek to understand whether \(m_2\) could indeed be a \(1.4\,M_\odot\) neutron star by performing a Bayesian analysis to extract the conditional probability distribution of its tidal deformability under this equal-mass assumption.

In this framework, we reduce the free parameters to a single tidal deformability value, \(\Lambda_2\). We define its prior as
\begin{equation}
\Pr(\Lambda_2) \sim \mathcal{U}(0, 2000).
\end{equation}
The likelihood for \(\Lambda_2\) is derived from the same KDE-based approach described earlier, but now evaluated under the condition that \(M_2 = 1.4\,M_\odot\). Specifically,
\begin{equation}
\mathcal{L}(\Lambda_2) \;=\; P\bigl(O \mid \Lambda_2,\, M_2 = 1.4\,M_\odot\bigr),
\end{equation}
where \(O\) denotes the observed data for GW170817.

The posterior distributions of $\Lambda_2$ reflect the tidal deformability of each neutron star under the condition that assuming they all $1.4 M_\odot$ masses.
we then could use bayes' theorem define the \lone as:

\begin{equation}
\begin{aligned}
P(\Lambda_{1.4}) \propto  \;&Pr(\Lambda_2)\cdot P(O \mid \Lambda_2, M_2 = 1.4 M_\odot)\\ &\cdot P(M_2 = 1.4 M_\odot \mid O)
\end{aligned}
\label{overall}
\end{equation}

To assess the reliability of the assumption that \(m_2\) has a mass exactly equal to \(1.4\,M_{\odot}\), we incorporate its likelihood factor $P(M_2 = 1.4 M_\odot \mid O)$ explicitly. This joint posterior distribution thus represents the inferred tidal deformability for a \(1.4\,M_{\odot}\) neutron star, based on GW170817 data, under the premise that one component of the binary indeed has this canonical mass. All Bayesian inferences in this study were conducted using the nested-sampling package UltraNest \citep{2021JOSS....6.3001B}. For each inference, we employed 50,000 live points to ensure convergence. All likelihood and computation algorithms are implemented in the CompactObject package \citep{compactobject}, an open-source, full-scope Bayesian inference framework developed by the author, specifically designed for neutron star physics.
\begin{figure}
	\centering
	\includegraphics[scale=0.5]{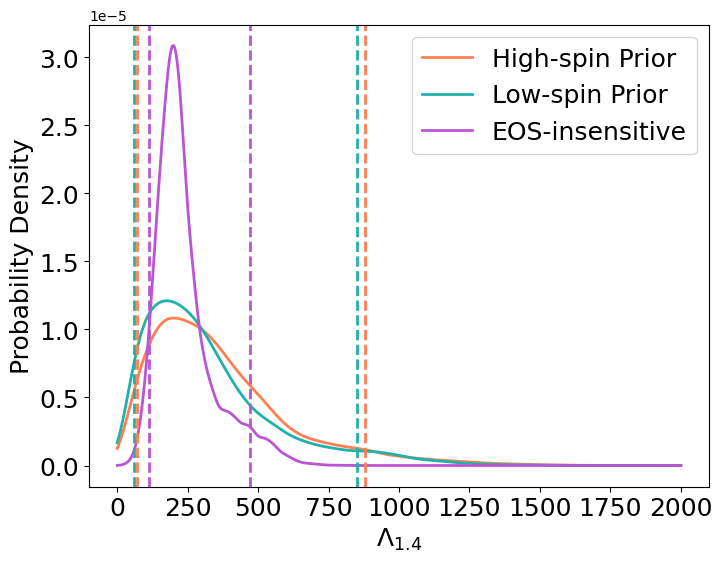}
	\caption{The 1-D distribution of 1.4~\msol{} star tidal deformablity \lone from three different data set: high-spin prior data set (red), low-spin prior data-set (green), EOS-insensitive  data set (purple): The dashed lines in each figure from left to right represent the quantiles of each distribution at 5\% quantile maximum probablity density peak value location and 90\% quantile. The density function is computed using KDE estimation. 
	}
	\label{EOS_independent}
\end{figure} 

\begin{table*}
\centering
\begin{tabular}{cccc}
\hline \hline
      \text{   GW170817 data-set    }&\text{$\Lambda_{1.4}$}&\text{\ron} \\ 
\hline
      \text{High-spin prior}    & $262.29_{-161.09}^{+917.82}$& $11.51_{-1.37}^{+2.56}$ km     \\
      \text{Low-spin prior} &  $276.02_{-167.31}^{+758.85}$  & $11.59_{-1.35}^{+2.23}$ km  \\

      \text{EOS-insensitive}    & $222.89_{-98.85}^{+420.33}$  & $11.26_{-0.83}^{+1.71}$ km     \\
\hline 
\hline
\end{tabular}
\caption{This table sumarize all the predictions for \lone from different scenario 1 with different data-sets: High-spin prior data, low-spin prior data and EOS-insensitive data,  the interpolated $R_{1.4}$ of derived from emperical relation $\Lambda\left(R_{1.4}\right)=2.88 \times 10^{-6}\left(R_{1.4} / \mathrm{km}\right)^{7.5}$ in \cite{Eemeli}. }
\label{R14_table}
\end{table*}
\subsection{Implemented dataset}
The GW170817 event marked the first detection of a double neutron star merger, offering a direct measurement of the tidal deformabilities of both constituent stars. In this work, we do not reanalyze the original strain data. Instead, we focus on directly extracting \(\Lambda_{1.4}\) from the publicly available posteriors of the masses and tidal deformabilities of the two neutron stars in GW170817. Below, we summarize the three data sets employed in our study.

The first two data sets are drawn from the original GW170817 discovery paper, based on a post-Newtonian (PN) waveform model and two spin priors:
\begin{itemize}
    \item \textbf{High-spin prior:} dimensionless spin \(\lvert \chi \rvert \leq 0.89\)
    \item \textbf{Low-spin prior:} dimensionless spin \(\lvert \chi \rvert \leq 0.05\)
\end{itemize}
These posterior samples are considered ``EOS-irrelevant'' or at least relatively free from specific EOS assumptions, since a PN waveform model does not rely strongly on detailed neutron-star structure inputs. Each posterior set contains joint probability distributions of \(\bigl(M_1, \Lambda_1\bigr)\) and \(\bigl(M_2, \Lambda_2\bigr)\).

The third data set derives from additional analyses that incorporate the ``EOS-insensitive'' universal relations proposed by ~\cite{PhysRevD.88.023007,Urbanec_2013,Yagi_2016,Yagi_2017,Chatziioannou_2018}. These universal relations apply to a broad range of EOS models and can constrain the parameter space more tightly than a purely PN-based approach, but they remain relatively model-agnostic. Specifically, the posterior is still considered EOS-insensitive because it does not assume one specific EOS; however, the universal-relation constraints narrow the allowed \(\Lambda\)-range.

For a complete description of these data sets and their posterior extraction procedures, we refer readers to \cite{Abbott_2017,Abbott_2018}. We emphasize that our focus is on the mass--tidal-deformability relation, although the full posteriors also include additional parameters describing the binary system. Incorporating these three data sets into the two scenarios described earlier (different masses vs.\ identical masses) enables a straightforward and robust estimation of \(\Lambda_{1.4}\).

{\subsection{Justification of the linear interpolation assumption}\label{sec:phys_motivation}
}
{Tidal deformability is given by
\begin{equation}
  \Lambda = \frac{2}{3}k_2\,\left(\frac{R\,c^2}{G\,M}\right)^5 
          \;=\; \frac{2}{3}k_2\,C^{-5},
  \label{eq:Lambda_basic}
\end{equation}
}
{
where $k_2$ is the dimensionless Love number and $C\equiv GM/(Rc^2)$ is the stellar compactness.  
{At \emph{fixed} $M$, $\Lambda$ scales trivially as $R^5$;} however, $C(m)$ contains the full microphysics of the EOS and may host non-analytic features such as phase transitions \citep[e.g.][]{Yagi_2017,Essick_2020}.  
{A common approach in the literature \citep{Yagi_2016a,De_2018,Abbott_2019,Zhao_2018,annala_2018} }is to Taylor-expand $\Lambda m^5$ about $m=1.4\,M_\odot$,
\begin{equation}
  \Lambda(m)\,m^5 = 
    \Lambda_{1.4}\,m_{1.4}^5 + 
    \Bigl.\frac{d(\Lambda m^5)}{dm}\Bigr|_{1.4}\!(m-1.4)
    + \mathcal{O}\!\bigl((m-1.4)^2\bigr),
  \label{eq:m5_expansion}
\end{equation}
}
{
which implicitly assumes $\Lambda\propto m^{-6}$ and that the slope $d(\Lambda m^5)/dm$ varies slowly.  This is the standard expansion method widely employed in the gravitational-wave community. (Hereafter, all references to the “standard” method pertain to this expansion.)}

{To test these premises we surveyed 200 state-of-the-art EOSs -- radomly generated from \texttt{CompactObject} package developed by the author \citep{compactobject}, using the Relativistic Mean Field theory model defined in \cite{Huang:2023grj}, keep the same prior range as the generating range. 
The Monte-Carlo bias test in Fig.~\ref{fig:bias_test} shows that within the viable GW170817 range, compare to the injected $\Lambda_{1.4}$ defined by these EOSs, and the linear interpolated model values, the data is distributed around unity with different mass gap that defined by the computation position of two randomly picking point along the EOS. This qualitatively show the agreement between our linear approximation and the injected value. Quantitatively, at the $\Lambda_{1.4}$ {smaller} than 1500 over the interval $1.2$–$1.6\,M_\odot$ the function $\Lambda(m)$ is nearly linear with a median fractional residual $|\delta_{\rm lin}|\approx10\%$ .}

{Because the current observational uncertainty on $\Lambda_{1.4}$ from GW170817 is $\gtrsim40\%$ \citep{Abbott_2017,Abbott_2018,Abbott_2019}, }adopting the simpler linear form in $\Lambda$ minimises higher-order truncation error while remaining agnostic about the exact $C(m)$ relation.

{
The Taylor series of $\Lambda m^{5}$ around $m=1.4\,M_\odot$ referred in \ref{eq:m5_expansion}
rests on two subtle assumptions:  
(\emph{i}) that, to leading order, $\Lambda\!\propto\! m^{-6}$, and  
(\emph{ii}) that the derivative ${\rm d}(\Lambda m^{5})/{\rm d}m$ varies
smoothly within the $\simeq0.3\,M_\odot$ window used for the expansion.
These conditions are satisfied when the compactness--mass relation
$C(m)=GM/(Rc^{2})$ is monotonic and analytic, as in purely nucleonic or
moderately soft equations of state (EOSs).
By contrast, a \emph{linear interpolation} of $\Lambda(m)$ imposed here, has no pre-defined
functional form on $C(m)$; it merely follows the local slope of the
empirical $(m,\Lambda)$ curve.
This linear interpolation approach therefore remains reliable in several physically
motivated situations where the ``standard'' expansion can break down:
{\begin{enumerate}
    \setlength\itemsep{2pt}
  \item Strong first--order phase transitions (e.g.\ onset of
        deconfined quark matter or hyperons) that introduce an
        inflection or kink near $1.4\,M_\odot$ \citep{Han_2019,Counsell2025};
  \item Twin--star configurations, in which $\Lambda(m)$ stays
        monotonic but $\Lambda m^{5}$ acquires sharp curvature \citep{landry2022prospectsconstrainingtwinstars,Tsaloukidis:2022rus,Christian_2019a,zhou2025};
  \item Rapid soft--stiff cross--overs driven by changes in the
        symmetry--energy slope $L$ \citep{hu2020,PhysRevC.102.045807,PhysRevC.104.015802}; and
  \item Rapid rotation, which violates the strict
        $\Lambda\propto C^{-5}$ scaling \citep{salinas2024assessingimpactuniformrotation,Yu_2024,PhysRevResearch.3.033129,Kr_ger_2021}.
\end{enumerate}}
For such EOSs the higher--order terms neglected in the
$\Lambda m^{5}$ series become non‐negligible, whereas a local linear
interpolation of $\Lambda(m)$ still captures the dominant trend.
These statements motivate our preference for the linear
scheme when seeking an EOS‐agnostic constraint on $\Lambda_{1.4}$.}

\section{Inference Result}
In this section, we present our estimates of \(\Lambda_{1.4}\) using the GW170817 inference data sets, as described under the two scenarios outlined above. For each scenario, we discuss results obtained from different sample selections and compare them against the standard analysis methods commonly found in the literature. Our primary goal is to highlight how our approach may complement or diverge from these established techniques.

\subsection{Scenario 1}

In the first scenario, we assume \(m_1 \neq m_2 \neq 1.4\text{ } M_{\odot}\), treating the two neutron stars in GW170817 as distinct objects, each with its own mass and tidal deformability. We then perform the sampling procedure described in Section~2, interpolating to estimate \(\Lambda_{1.4}\), the tidal deformability of a \(1.4\,M_\odot\) neutron star.

Figure~\ref{fig:highlow_compare} shows the one-dimensional posterior distributions of \(\Lambda_{1.4}\) derived from the high-spin and low-spin prior analyses. 
\begin{itemize}
    \item \textbf{High-spin prior} (\(\lvert \chi \rvert \leq 0.89\)): We obtain \(\Lambda_{1.4} = 262.29_{-161.09}^{+917.82}\). Comparing to the standard expansion method, which places the 95\% upper limit at \(\Lambda_{1.4} \leq 1400\), our result is broadly consistent within uncertainties.  
    \item \textbf{Low-spin prior} (\(\lvert \chi \rvert \leq 0.05\)): Here, the posterior is tighter than in the high-spin case, consistent with the narrower prior range on the spin. Our upper bound for \(\Lambda_{1.4}\) is \(\approx 1034.87\), whereas the standard expansion method yields \(\Lambda_{1.4} \leq 970\). Despite this slight difference, both the peak values and the lower bounds remain largely consistent across the two approaches, reflecting similar quantiles for the low-spin and high-spin analyses.
\end{itemize}

\noindent
\textbf{EOS-Insensitive (Universal-Relations) Dataset.} 
In \cite{Abbott_2018}, an additional analysis adopted EOS-insensitive universal relations to constrain \(\Lambda_1\) and \(\Lambda_2\). Although these relations are not purely EOS-irrelevant, they are generally considered model-agnostic. The analysis also imposes \(\Lambda_1 \leq \Lambda_2\), reflecting the expectation that the more massive star has a smaller tidal deformability for most viable EOS models. We adopt the same condition here.

Figure~\ref{EOS_independent} shows our posterior distribution for \(\Lambda_{1.4}\) in this EOS-insensitive analysis, yielding
\[
\Lambda_{1.4} = 222.89_{-98.85}^{+420.33}.
\]
By comparison, the standard expansion approach gives \(\Lambda_{1.4} = 190_{-120}^{+390}\). These results are largely consistent within their respective uncertainties. Our median is slightly higher than the standard expansion value, though the lower bound is smaller, and the overall 5\%--95\% interval is similar. Because our method interpolates directly from the \(\Lambda_1\)--\(\Lambda_2\) posteriors, these uncertainties are reasonable and expected. This agreement between our model-independent approach and the standard expansion method implies that higher-order corrections in the linear expansion are not critical at the current level of observational precision. However, as gravitational-wave detectors improve in sensitivity, such corrections might become more important, and a fully data-driven approach like ours could offer a more robust inference free from linearization assumptions.

Finally, \cite{Abbott_2018} also provides results using parametrized EOS models. While our method could be applied to these data sets, our goal is to maintain an EOS-independent framework. Nonetheless, we expect consistency with the EOS-insensitive analysis, as they typically predict similar distributions for \(\Lambda_1\) and \(\Lambda_2\).

\subsection{Scenario 2}
In Scenario~1, we specifically select samples satisfying
$
M_1 < 1.4\,M_\odot < M_2,
$
since the posterior mass range for \(M_2\) includes \(1.4\,M_\odot\). To supplement this, we now consider the possibility that \(M_2\) is \emph{exactly} \(1.4\,M_\odot\). From a formation standpoint, the likelihood of a star having precisely \(1.4\,M_\odot\) is small; however, within measurement uncertainties, there is a nonzero probability that the star’s mass is exactly \(1.4\,M_\odot\). We can quantify that probability based on the posterior distribution.

Figure~3 shows the \(\Lambda_{1.4}\) probability distributions for all three data sets (high-spin, low-spin, and EOS-insensitive priors) under the assumption \(M_2 = 1.4\,M_\odot\). In the high-spin prior case, we obtain \(\Lambda_{1.4} = 200.05_{-129.09}^{+678.72}\), exhibiting the broadest 90\% range and mirroring the behavior observed in Scenario~1. The low-spin prior yields \(\Lambda_{1.4} = 175.54_{-116.67}^{+675.36}\), reflecting a somewhat narrower 90\% interval due to more restrictive assumptions on the star’s spin. Finally, when the EOS-insensitive universal-relation constraint is imposed, the result becomes \(\Lambda_{1.4} = 198.54_{-85.66}^{+274.13}\), displaying a decreased central value and a tighter 90\% interval that aligns with the trend seen in Scenario~1.

Notably, these values of \(\Lambda_{1.4}\) are generally lower and have tighter 90\% intervals compared to the results in Scenario~1. However, their \emph{overall} probability densities are significantly suppressed by the factor
$
P\bigl(M_2 = 1.4\,M_\odot \mid O\bigr),
$
which quantifies how likely \(M_2\) is exactly \(1.4\,M_\odot\) given the observational data \(O\). Because this factor is small, it contributes negligibly to the final \(\Lambda_{1.4}\) inference when combining all samples. 

On the other hand, this finding underscores that if future measurements achieve higher precision on the component masses of a binary neutron star system, the scenario of an exactly \(1.4\,M_\odot\) object could yield a strikingly tight \(\Lambda_{1.4}\) estimate. In the present case, however, the current mass uncertainties make the probability of \(M_2 = 1.4\,M_\odot\) too small to significantly affect our overall inference of \(\Lambda_{1.4}\).

\section{Multimessenger constraint on $R_{1.4}$ and $\Lambda_{1.4}$}
Table~\ref{R14_table} summarizes all of our Scenario~1 results for the three GW170817 data sets. From each tidal deformability inference, we also derive the corresponding \(R_{1.4}\) constraints.

\paragraph{Combining GW170817 with X-ray timing Observations.}
As discussed in \cite{Huang_2025}, the methodology can be extended to more than two sources. Here, since we have a purely GW-based constraint on \(\Lambda_{1.4}\) from GW170817, we can combine it with the constraints from the NASA’s Neutron Star Interior Composition Explorer (NICER) \citep{gendreau2016neutron} X-ray observations of PSR J0030+0451 (J0030) \citep{Miller_2019,Riley_2019,Vinciguerra_2024} and PSR J0437$-$4715 (J0437) \citep{Choudhury_2024} to obtain a joint multimessenger limit on \(\Lambda_{1.4}\) and \(R_{1.4}\). Because the GW and X-ray results are statistically independent---and both are largely EOS-agnostic---we can multiply their respective posterior distributions to obtain a combined EOS-insensitive constraint.

From the NICER side, different model assumptions regarding J0030’s hotspot geometry can influence the mass-radius inference. We refer the reader to \cite{Huang_2025,Vinciguerra_2024} for a detailed discussion of these model choices. Following \cite{Huang_2025}, we adopt the ST+PDT model for J0030, which is argued to offer a robust reference scenario. For the GW170817 data, we select the EOS-insensitive universal-relations data set as our representative best-estimate result.

Multiplying these two posteriors (GW-based and NICER-based) gives:
\[
R_{1.4} = 11.53_{-0.88}^{+0.89} \,\text{km}, 
\quad
\Lambda_{1.4} = 265.18_{-104.38}^{+237.88}.
\]
These values do not incorporate the PSR J0740+6620 constraint \citep{salmi2024radiushighmasspulsar,2024ApJ...974..295D}, because that source’s mass is around \(2.0\,M_\odot\), which would require additional considerations when interpolating \(\Lambda(M)\). Consequently, our final constraints remain somewhat broad. Nonetheless, this is one of the first EOS-independent multimessenger constraints on \(\Lambda_{1.4}\), making it a valuable benchmark for understanding neutron-star interiors and nuclear physics in the beyond-saturation-density regime.


\section{Discussion}
Constraining \(\Lambda_{1.4}\) and \(\Lambda(R_{1.4})\) has been a long-standing challenge. Many approaches, including nonparametric methods, have been proposed to extract \(\Lambda_{1.4}\) from GW and NICER observations. Our study aims to offer a purely data-driven perspective. In contrast, many other observational constraints favor a central value of \(\Lambda_{1.4}\) around 400, whereas our results at the 95\% credibility level indicate \(\Lambda_{1.4} < 503.06\). This upper limit is close to the central value in some previous studies, largely because of including PSR J0740+6620 observations, whose radius is comparable to those of J0030 and J0437. The relatively large radius implied by J0740+6620 favors a stiffer EOS, thereby pushing \(\Lambda_{1.4}\) to higher values. However, this conclusion also depends heavily on the choice of EOS framework. Since our method is nearly EOS-independent, and the true neutron-star EOS remains uncertain, having a purely model-agnostic result serves as a valuable reference.

Cross-comparisons between our GW-based constraints on \(\Lambda_{1.4}\) and those derived from NICER data in \cite{Huang_2025} reveal strong agreement in both central values and uncertainties. We therefore see no compelling need for introducing a twin-star assumption in this mass range (i.e., two distinct radii at the same mass), given that \(R_{1.4}\) inferred from two different observational approaches remains in good agreement. Furthermore, by comparing with phase transition study in \cite{Huang:2025vfl}, we note that our inferred \(R_{1.4}\) from the $\Lambda_{1.4}$ result is \(\sim1\) km smaller, suggesting that a phase transition at this mass range is not strongly favored, consistent with the presented Bayesian evidence in \cite{Huang:2025vfl}.

Nuclear experiments such as P-REX and C-REX also constrain \(\Lambda_{1.4}\) and \(R_{1.4}\) \citep{PREX:2021umo}, although their results depend on which EOS model is adopted \citep{Reinhard21,Reinhard:2022inh,Mondal:2022cva}. Our multimessenger constraint appears more consistent with the C-REX outcome than with P-REX.(see \cite{PREX:2021umo}) Given the tension between C-REX and P-REX measurements, our EOS-independent astrophysical constraints may help guide further resolution of these experimental discrepancies.

Finally, the result from Scenario~2, as discussed previously, is not treated here as a definitive inference, given its low overall posterior probability density. Nevertheless, this method will become increasingly valuable if future observations provide more precise mass measurements, particularly for sources close to \(1.4\,M_\odot\). With multiple binary neutron-star merger events similar to GW170817, and additional systems near the canonical mass, this approach could yield strong constraints. Under current observational limitations, however, its impact remains negligible due to the small probability associated with having an exactly \(1.4\,M_\odot\) star.

\section{Conclusion}
In this work, we have introduced a purely data-driven framework for constraining the tidal deformability \(\Lambda_{1.4}\) of a canonical \(1.4\,M_{\odot}\) neutron star, applying our method to gravitational-wave (GW) observations from GW170817 and complementary X-ray measurements from NICER. By interpolating directly in \((M,\Lambda)\) space, our approach bypasses explicit assumptions about the neutron-star equation of state (EOS). We demonstrated that when selecting samples satisfying \(M_1 < 1.4\,M_\odot < M_2\), the inferred \(\Lambda_{1.4}\) values from high-spin, low-spin, and EOS-insensitive data sets remain broadly consistent with more traditional expansion-based methods, suggesting that higher-order corrections in the linear expansion do not significantly affect inferences at the current level of observational precision. We provided the posterior value for $\Lambda_{1.4} = 222.89_{-98.85}^{+420.33}$ based on EOS-insensitive data set.

Another component of our analysis explored the possibility of a precisely \(1.4\,M_\odot\) star in the binary, revealing that the resulting \(\Lambda_{1.4}\) posterior distribution can be notably narrower but is heavily suppressed by the low probability of having exactly \(1.4\,M_\odot\). While these mass constraints do not substantially influence the overall inference at present, they highlight an avenue for future work, especially as more sensitive detectors and additional binary neutron-star mergers yield better mass measurements.

We further combined our GW-based \(\Lambda_{1.4}\) inferences with NICER observations of PSR J0030+0451 and PSR J0437$-$4715, reinforcing the advantages of a multimessenger strategy. Because the GW and X-ray posteriors from this work and \cite{Huang_2025} are statistically independent and largely EOS-agnostic, their product delivers a more robust and unified constraint on \(\Lambda_{1.4} = 265.18_{-104.38}^{+237.88}\) and \(R_{1.4} = 11.53_{-0.88}^{+0.89} \,\text{km}\). The resulting joint constraints still align well with terrestrial experiments and other astrophysical observations. This consistency underscores how EOS-independent measurements can help interpret the often divergent outcomes of nuclear-physics experiments such as P-REX and C-REX, which may rely on their own EOS modeling assumptions.

Overall, our findings confirm that a straightforward, interpolation-based method can achieve results comparable to established expansion or parametric-EOS methods when data uncertainties are relatively large. As future GW observatories and NICER-like missions refine mass and tidal deformability measurements, it may become increasingly important to avoid linearization assumptions in current standard approach that risk systematic biases. The demonstrated compatibility between GW and X-ray constraints points to the promise of expanding this multimessenger approach, particularly once multiple high-precision measurements of binary neutron-star mergers and pulsar masses become available. 

\section{Software and third party data repository citations} \label{sec:cite}

\textit{CompactObject package: An full-scope open-source Bayesian inference framework especially designed for Neutron star physics} {Zenodo repository of CompactObject package: \url{https://zenodo.org/records/14181695}} for Version 2.0.0, Github: \url{https://github.com/ChunHuangPhy/CompactObject}, documentation: \url{https://chunhuangphy.github.io/CompactObject/}. \textit{UltraNest}: \cite{2021JOSS....6.3001B}, \url{https://github.com/JohannesBuchner/UltraNest}.

\section{acknowledgments}
C.H. acknowledges support from NASA grant 80NSSC24K1095. The author extends gratitude to Mark Alford, Alexander Chen, Ryan O'Connor and Chuyi Deng for insightful discussion.




\bibliography{sample631}{}
\bibliographystyle{aasjournal}



\end{document}